# The Influence of Two Strong Electromagnetic Waves on Electron

K.V. Ivanyan*

M.V. Lomonosov Moscow State University, Moscow 119991, Russia

We consider the motion of a nonrelativistic electron in the field of two strong monochromatic light waves propagating counter to each other. The wave function of the electron is obtained by using a quasiclassical approximation and perturbation theory.

**INTRODUCTION**

The motion of an electron in the field of a monochromatic light wave is described by the well-known volkov function. Exact solutions of the relativistic wave equations were obtained in [1,2] for the motion of an electron in certain cases of plane-wave fields. The motion of an electron in the field of two light waves propagating counter to each other (standing wave), however, cannot be solved exactly.

Electron diffraction by a standing wave was considered by perturbation theory in [3]. The electron channeling was investigated in in intense standing light wave in [4-6]. Other schemes for FEL were considered in [7-57].

We consider in the present paper a quasiclassical approximation for the description of the motion of a nonrelativistic electron in the field of pump waves. The wave functions of the nonrelativistic electrons in the field of pump waves (standing wave) are obtained using a quasiclassical approximation. The corrections to these wave functions, necessitated by the rescattering effect (the electron absorbs a photo from one wave and gives it up to the other by induced emission) are obtained by perturbation theory. The electron energy is not altered by the rescattering, but the momentum is changed by an amount $\mathbf{p'} = \mathbf{p} + 2\hbar\mathbf{k}$, where $\mathbf{p'}$ and $\mathbf{p}$ are the electron momenta in the initial and final states, respectively, while $\mathbf{k}$ is the pump-wave momentum. The electron momentum change can be accompanied by emission in the IR band.

-----------------

*k.ivanyan@yandex.com

## GENERAL RELATIONS

The Schrodinger equation for an electron in the field of two strong electromagnetic waves is of the form

$$i\hbar \frac{\partial \Psi}{\partial t} = -\frac{\hbar^2}{2m}\Delta\Psi(\mathbf{r},t) + i\hbar \frac{e}{mc}[\mathbf{A}_1(\varphi_1) + \mathbf{A}_2(\varphi_2)]\nabla\Psi(\mathbf{r},t)$$
$$+ \frac{e^2}{2mc^2}[\mathbf{A}_1^2(\varphi_1) + \mathbf{A}_2^2(\varphi_2)]\Psi(\mathbf{r},t) + \frac{e^2}{mc^2}\mathbf{A}_1(\varphi_1)\mathbf{A}_2(\varphi_2)\Psi(\mathbf{r},t), \qquad (1)$$

where $\mathbf{A}_1(\varphi_1)$ and $\mathbf{A}_2(\varphi_2)$ are the vector potentials of the electromagnetic waves, respectively; where $\varphi_1 = \omega_1 t - \mathbf{k}_1\mathbf{r}_1$, $\varphi_2 = \omega_2 t - \mathbf{k}_2\mathbf{r}$ ($\omega_1, \omega_2$ are respectively the frequencies of the first and second electromagnetic waves, and $\mathbf{k}_1$ and $\mathbf{k}_2$ are the wave vectors of these waves); $m$ is the electron mass.

We seek the solution of Eq. (1) in the form

$$\Psi(\mathbf{r},t) = \frac{1}{\sqrt{V}}\exp\left[i\frac{\mathbf{pr}}{\hbar} - i\frac{Et}{\hbar}\right]\exp\left[i\frac{S_1(\varphi_1)}{\hbar} + i\frac{S_1(\varphi_1)}{\hbar}\right]\Psi_{int}(\mathbf{r},t), \qquad (2)$$

where V is the normalization volume.

Here p and $E$ are the momentum and kinetic energy of the electron:

$$S_1(\varphi_1) = \frac{1}{m\omega_1}\int\left[\frac{e}{c}\mathbf{A}_1(\varphi_1)\mathbf{p} - \frac{e^2}{2c^2}\mathbf{A}_1^2(\varphi_1)\right]d\varphi_1, \qquad (3)$$

$$S_2(\varphi_2) = \frac{1}{m\omega_2}\int\left[\frac{e}{c}\mathbf{A}_2(\varphi_2)\mathbf{p} - \frac{e^2}{2c^2}\mathbf{A}_2^2(\varphi_2)\right]d\varphi_2. \qquad (4)$$

Substitution of relations (2), (3), and (4) in (1) leads to an equation for $\Psi_{int}(\mathbf{r},t)$:

$$i\hbar\frac{\partial \Psi_{int}(\mathbf{r},t)}{\partial t} = -\frac{\hbar^2}{2m}\Delta\Psi_{int}(\mathbf{r},t) - \frac{i\hbar}{m}\left\{\mathbf{p} - \mathbf{k}_1\frac{dS_1(\varphi_1)}{d\varphi_1} - \mathbf{k}_2\frac{dS_2(\varphi_2)}{d\varphi_2}\right.$$
$$\left. -\frac{e}{c}[\mathbf{A}_1(\varphi_1) + \mathbf{A}_2(\varphi_2)]\right\}\nabla\Psi_{int}(\mathbf{r},t) + \left[\frac{e^2}{mc^2}\mathbf{A}_1(\varphi_1)\mathbf{A}_2(\varphi_2) + \frac{1}{m}\mathbf{k}_1\mathbf{k}_2\frac{dS_1(\varphi_1)}{d\varphi_1}\frac{dS_2(\varphi_2)}{d\varphi_2}\right. \qquad (5)$$
$$\left. +\frac{e}{mc}\mathbf{A}_1\mathbf{k}_2\frac{dS_2(\varphi_2)}{d\varphi_2} + \frac{e}{mc}\mathbf{A}_2\mathbf{k}_1\frac{dS_1(\varphi_2)}{d\varphi_1}\right]\Psi_{int}(\mathbf{r},t).$$

We can leave out of (5) the terms

$$\mathbf{k}_1 \frac{dS_1(\varphi_1)}{d\varphi_1}, \qquad \mathbf{k}_2 \frac{dS_2(\varphi_2)}{d\varphi_2}, \qquad -\frac{e}{c}[\mathbf{A}_1(\varphi_1) + \mathbf{A}_2(\varphi_2)],$$

$$\frac{1}{m}\mathbf{k}_1\mathbf{k}_2 \frac{dS_1(\varphi_1)}{d\varphi_1}\frac{dS_2(\varphi_2)}{d\varphi_2}, \qquad \frac{e}{mc}\mathbf{A}_1\mathbf{k}_2 \frac{dS_2(\varphi_2)}{d\varphi_2}, \qquad \frac{e}{mc}\mathbf{A}_2\mathbf{k}_1 \frac{dS_1(\varphi_2)}{d\varphi_1}$$

if the inequalities" $cp\cos\theta \gg e\varepsilon_i \lambdabar_i$ ($i = 1,2$) and $v/c \ll 1$.

Equation (5) takes then the form

$$i\hbar \frac{\partial \Psi_{int}(\mathbf{r},t)}{\partial t} = -\frac{\hbar^2}{2m}\Delta\Psi_{int}(\mathbf{r},t) - \frac{i\hbar}{m}\mathbf{p}\nabla\Psi_{int}(\mathbf{r},t) + \frac{e^2}{mc^2}\mathbf{A}_1(\varphi_1)\mathbf{A}_2(\varphi_2)\Psi_{int}(\mathbf{r},t). \qquad (6)$$

We consider for the sake of argument linearly polarized electromagnetic waves

$$\mathbf{A}_1(\varphi_1) = \mathbf{A}_1 \sin\varphi_1, \qquad (7)$$

$$\mathbf{A}_2(\varphi_2) = \mathbf{A}_2 \sin\varphi_2. \qquad (8)$$

Substituting (7) and (8) in (6), we get

$$i\hbar \frac{\partial \Psi_{int}(\mathbf{r},t)}{\partial t} = -\frac{\hbar^2}{2m}\Delta\Psi_{int}(\mathbf{r},t) - \frac{i\hbar}{m}\mathbf{p}\nabla\Psi_{int}(\mathbf{r},t) \\ + \frac{e^2}{2mc^2}\mathbf{A}_1\mathbf{A}_2[\cos(\varphi_1 - \varphi_2) - \cos(\varphi_1 + \varphi_2)]\Psi_{int}(\mathbf{r},t). \qquad (9)$$

It is natural to seek the solution of (9) in the form

$$\Psi_{int}(\mathbf{r},t) = \Psi_{int}^{(0)}(\mathbf{r},t) e^{iS_3(\mathbf{r},t)/\hbar}, \qquad (10)$$

where, if the inequalities $pc \gg e\varepsilon_1 \lambdabar_i$ ($i = 1,2$), we have the following equations for $S_3(\mathbf{r},t)$ and $\Psi_{int}^{(0)}(\mathbf{r},t)$:

$$i\hbar \frac{\partial S_3(\mathbf{r},t)}{\partial t} = -\frac{i\hbar}{2m}\Delta S_3(\mathbf{r},t) + \frac{1}{2m}[\nabla S_3(\mathbf{r},t)]^2 + \frac{\mathbf{p}}{m}\nabla S_3(\mathbf{r},t) - \frac{e^2\mathbf{A}_1\mathbf{A}_2}{2mc^2}\cos(\varphi_1 + \varphi_2) \qquad (11)$$

$$i\hbar \frac{\partial \Psi_{int}^{(0)}(\mathbf{r},t)}{\partial t} = -\frac{\hbar^2}{2m}\Delta\Psi_{int}^{(0)}(\mathbf{r},t) - i\hbar\frac{\mathbf{p}}{m}\nabla\Psi_{int}^{(0)}(\mathbf{r},t) + \frac{e^2\mathbf{A}_1\mathbf{A}_2}{2mc^2}\cos(\varphi_1 - \varphi_2)\Psi_{int}^{(0)}(\mathbf{r},t). \qquad (12)$$

It follows from the forms of (11) and (12) that the solutions of these equations should depend on $\varphi_1 + \varphi_2$ and $\varphi_1 - \varphi_2$, respectively. Neglecting the first, second, and third terms in the right-hand side of (11), we obtain the solution

$$S_3(\mathbf{r},t) = \frac{e^2 A_1 A_2}{2mc^2(\omega_1 + \omega_2)} \sin(\varphi_1 + \varphi_2) \tag{13}$$

Neglect of the first, second, and third terms in relation (11) is valid subject to the inequalities

$$\hbar(\omega_1 + \omega_2)/2mc^2 \ll 1, \quad e^2 A_1 A_2 /(2mc^2)^2 \ll 1, \quad v \ll c.$$

We seek the solution of (12) by introducing first a new variable $z = \varphi_1 - \varphi_2$; we can then rewrite this equation in the form

$$\frac{\hbar^2(\mathbf{k}_2 - \mathbf{k}_2)^2}{2m} \frac{d^2 \Psi_{int}^{(0)}(z)}{dz^2} + i\hbar[\omega_1 - \omega_2 + \mathbf{v}(\mathbf{k}_2 - \mathbf{k}_1)] \frac{d\Psi_{int}^{(0)}(z)}{dz} - \frac{e^2 A_1 A_2}{2mc^2} \cos(z) \Psi_{int}^{(0)}(z) \tag{14}$$

or

$$\frac{\partial^2 \Psi_{int}^{(0)}(z)}{\partial z^2} + i \frac{2m[\omega_1 - \omega_2 + \mathbf{v}(\mathbf{k}_2 - \mathbf{k}_1)]}{\hbar(\mathbf{k}_2 - \mathbf{k}_1)^2} \frac{d\Psi_{int}^{(0)}(z)}{dz} - \frac{e^2 A_1 A_2}{(\hbar c)^2 (\mathbf{k}_2 - \mathbf{k}_1)^2} \cos z \, \Psi_{int}^{(0)}(z) = 0, \tag{15}$$

where v is the initial velocity of the electron.

We eliminate from (15) the term with the first derivative with respect to z, by making the change of variables

$$\Psi_{int}^{(0)}(z) = P(z) u(z), \tag{16}$$

where

$$P(z) = \exp\left\{-i \frac{[\omega_1 - \omega_2 + \mathbf{v}(\mathbf{k}_2 - \mathbf{k}_1)]m}{\hbar(\mathbf{k}_2 - \mathbf{k}_1)^2} z\right\}. \tag{17}$$

We have for *u* (z) the equation

$$u''(z) + \left\{\left[\frac{\omega_1 - \omega_2 + \mathbf{v}(\mathbf{k}_2 - \mathbf{k}_1)}{\hbar(\mathbf{k}_2 - \mathbf{k}_1)^2} - m\right]^2 - \frac{e^2 A_1 A_2}{(\hbar c)^2 (\mathbf{k}_2 - \mathbf{k}_1)^2} \cos z\right\} u(z) = 0. \tag{18}$$

We shall solve (18) by perturbation theory. We represent $u(z)$ in the form
$$u(z) = u_0(z) + u_1(z), \tag{19}$$
where $u_0(z)$ satisfies the equation

$$u_0''(z) + \left[\frac{\omega_1 - \omega_2 + \mathbf{v}(\mathbf{k}_2 - \mathbf{k}_1)}{\hbar(\mathbf{k}_2 - \mathbf{k}_1)^2} - m\right]^2 u_0(z) = 0, \tag{19'}$$

and $u_1(z)$ the equation

$$u_1''(z) + \left[\frac{\omega_1 - \omega_2 + \mathbf{v}(\mathbf{k}_2 - \mathbf{k}_1)}{\hbar(\mathbf{k}_2 - \mathbf{k}_1)^2} - m\right]^2 u_1(z) = \frac{e^2 \mathbf{A}_1 \mathbf{A}_2}{(\hbar c)^2 (\mathbf{k}_2 - \mathbf{k}_1)^2} \cos z \, u_0(z) = 0. \tag{19''}$$

The solutions of (19') and (19'') are

$$\begin{aligned} u_0(z) &= \exp\left[i \frac{\omega_1 - \omega_2 + \mathbf{v}(\mathbf{k}_2 - \mathbf{k}_1)}{\hbar(\mathbf{k}_2 - \mathbf{k}_1)^2} mz\right], \\ u_1(z) &= -\frac{e^2 \mathbf{A}_1 \mathbf{A}_2 \sin z}{2mc^2[\hbar\mathbf{v}(\mathbf{k}_2 - \mathbf{k}_1) + \hbar(\omega_1 - \omega_2)]} \exp\left[i \frac{\omega_1 - \omega_2 + \mathbf{v}(\mathbf{k}_2 - \mathbf{k}_1)}{\hbar(\mathbf{k}_2 - \mathbf{k}_1)^2} mz\right]. \end{aligned} \tag{20}$$

To obtain (20) it was necessary to meet the conditions

$$\begin{aligned} &\left|\frac{\omega_1 - \omega_2 + \mathbf{v}(\mathbf{k}_2 - \mathbf{k}_1)}{\hbar(\mathbf{k}_2 - \mathbf{k}_1)^2} m\right| \gg 1 \\ &\left|\frac{\omega_1 - \omega_2 + \mathbf{v}(\mathbf{k}_2 - \mathbf{k}_1)}{(\mathbf{k}_2 - \mathbf{k}_1)^2} m\right| \gg \frac{|e^2 \mathbf{A}_1 \mathbf{A}_2|^{1/2}}{|c(\mathbf{k}_2 - \mathbf{k}_1)|}, \\ &\left|\frac{e^2 \mathbf{A}_1 \mathbf{A}_2}{2mc^2[\hbar\mathbf{v}(\mathbf{k}_2 - \mathbf{k}_1) + \hbar(\omega_1 - \omega_2)]}\right| \ll 1. \end{aligned} \tag{20'}$$

The last condition of (20') is violated as $\hbar \to 0$.

Taking (29) and (20) into account, we have

$$\Psi_{int}^{(0)}(z) = 1 - i \frac{e^2 \mathbf{A}_1 \mathbf{A}_2}{\{2mc^2[\mathbf{v}(\mathbf{k}_2 - \mathbf{k}_1) + (\omega_1 - \omega_2)]\}} \sin z. \tag{20''}$$

Substituting (10), (13), and (20'') in (2) and taking (3) and (4) into account, we obtain the wave function of the nonrelativistic electron in the field of two strong electromagnetic waves at an arbitrary arrangement of the vectors $\mathbf{k}_1, \mathbf{k}_2$ and $\mathbf{p}$. We shall, however, be interested hereafter only in a situation in which $\mathbf{k}_1$ and $\mathbf{k}_2$ are collinear and form a standing wave.

In this case the wave function takes the form

$$\Psi_i(z) = \Psi_i^{(0)}\left\{1 - \frac{e^2 \mathbf{A}_1 \mathbf{A}_2}{\hbar[\mathbf{v}(\mathbf{k}_2 - \mathbf{k}_1) + (\omega_1 - \omega_2)]} \sin z\right\}. \tag{21}$$

where $\Psi_i^{(0)}$ are basis functions defined by the relation

$$\begin{aligned} \Psi_i^{(0)} &= \frac{1}{\sqrt{V}} \exp\left[i \frac{\mathbf{pr}}{\hbar} - i \frac{Et}{\hbar}\right] \exp\left[-i \frac{e\mathbf{A}_1 \mathbf{p}}{mc\hbar\omega_1} \cos\varphi_1 - i \frac{e^2 A_1^2}{4mc^2\hbar\omega_1} \varphi_1 \right. \\ &\quad + \frac{e^2 A_1^2}{8mc^2\hbar\omega_1} \sin 2\varphi_1 - i \frac{e\mathbf{A}_2 \mathbf{p}}{mc\hbar\omega_2} \cos\varphi_2 - \frac{e^2 A_2^2}{4mc^2\hbar\omega_2} \varphi_2 \\ &\quad \left. + \frac{e^2 A_2^2}{8mc^2\hbar\omega_2} \sin 2\varphi_2 + \frac{e^2 \mathbf{A}_1 \mathbf{A}_2}{2mc^2\hbar(\omega_1 + \omega_2)} \sin(\varphi_1 + \varphi_2)\right]. \end{aligned} \tag{22}$$

It can be easily shown by direct calculation that the $\Psi_i^{(0)}$ satisfy the orthogonality relation (if $\mathbf{k}_1$ and $\mathbf{k}_2$ are collinear or form a standing wave)

$$\int \Psi_i^{(0)} \Psi_f^{(0)*} d\mathbf{r} = (2\pi)^3 \delta(\mathbf{p}-\mathbf{p}'),$$

where $\Psi_i^{(0)*}$ is obtained from $\Psi_i^{(0)}$ by replacing the exponents $\mathbf{p}$ by $\mathbf{p}'$. The expression (22) for the wave function of the nonrelativistic electron coincides in the case of two collinear electromagnetic waves with the wave function obtained in [1] in the nonrelativistic limit.